\documentclass[twocolumn,twocolappendix]{aastex7}

\usepackage{graphicx}		
\usepackage{url}
\usepackage{epstopdf}
\usepackage{color}
\usepackage{textcomp}
\usepackage{gensymb}
\usepackage{multirow}
\usepackage{ragged2e}
\usepackage{hyperref}
\usepackage{tikz}
\usepackage{orcidlink}
\usepackage{url}
\usepackage{amsmath}
\usepackage{booktabs}
\usepackage{comment}
\usepackage{afterpage}
\usepackage{mathtools}
\usepackage{tabularx}
\usepackage{bold-extra}
\usepackage{xspace}
\usepackage{relsize}
\usepackage[utf8]{inputenc}
\usepackage{newunicodechar}
\usepackage[encapsulated]{CJK}
\usepackage{ucs}
\usepackage{soul}
\usepackage[T1]{fontenc}
\usepackage{nicefrac}
\usepackage{float}
\usepackage{gensymb}

\newcommand{\kharma}[0]{{KHARMA}\xspace}
\newcommand{\ipole}[0]{{\tt{ipole}}\xspace}
\newcommand{\igrmonty}[0]{{\tt{igrmonty}}\xspace}
\newcommand{\harm}[0]{{\tt{harm}}\xspace}
\newcommand{\grim}[0]{{\tt{grim}}\xspace}
\def\sgra{Sgr~A$^*$\xspace}
\def\m87{M87$^*$\xspace}

\begin{document}

\title{Electromagnetic Observables of Weakly Collisional Black Hole Accretion}

\author{Vedant Dhruv\,\orcidlink{0000-0001-6765-877X}}
\affiliation{Department of Physics, University of Illinois, 1110 West Green St., Urbana, IL 61801, USA}
\affiliation{Illinois Center for Advanced Study of the Universe, 1110 West Green Street, Urbana, IL 61801, USA}
\email{vdhruv2@illinois.edu}

\author{Ben Prather\,\orcidlink{0000-0002-0393-7734}}
\affiliation{CCS-2, Los Alamos National Laboratory, P.O. Box 1663, Los Alamos, NM 87545, USA}
\email{bprather@lanl.gov}

\author{Mani Chandra\,\orcidlink{0000-0002-3927-2850}}
\affiliation{nOhm Devices, Inc., Cambridge, MA, USA}
\email{manic@nohm-devices.com}

\author{Abhishek V. Joshi\,\orcidlink{0000-0002-2514-5965}}
\affiliation{Department of Physics, University of Illinois, 1110 West Green St., Urbana, IL 61801, USA}
\affiliation{Illinois Center for Advanced Study of the Universe, 1110 West Green Street, Urbana, IL 61801, USA}
\email{avjoshi2@illinois.edu}

\author{Charles F. Gammie\,\orcidlink{0000-0001-7451-8935}}
\affiliation{Department of Physics, University of Illinois, 1110 West Green St., Urbana, IL 61801, USA}
\affiliation{Illinois Center for Advanced Study of the Universe, 1110 West Green Street, Urbana, IL 61801, USA}
\affiliation{Department of Astronomy, University of Illinois, 1002 West Green St., Urbana, IL 61801, USA}
\email{gammie@illinois.edu}

\begin{abstract}

The black holes in the Event Horizon Telescope sources Messier 87* and Sagittarius A* (\sgra) are embedded in a hot, collisionless plasma that is fully described in kinetic theory yet is usually modeled as an ideal, magnetized fluid. In this Letter, we present results from a new set of weakly collisional fluid simulations in which leading order kinetic effects are modeled as viscosity and heat conduction. Consistent with earlier, lower-resolution studies, we find that overall flow dynamics remain very similar between ideal and non-ideal models. For the first time, we synthesize images and spectra of \sgra from weakly collisional models---assuming an isotropic, thermal population of electrons---and find that these remain largely indistinguishable from ideal fluid predictions. However, most weakly collisional models exhibit lower light curve variability, with all magnetically dominated models showing a small but systematic decrease in variability. 
    
\end{abstract}

\section{Introduction}
\label{sec:introduction}

In recent years, the Event Horizon Telescope (EHT) has produced high-resolution radio images ($\lambda\sim$1.3mm) of the supermassive black hole at the center of the Milky Way \cite{SgrAPaperI, SgrAPaperII, SgrAPaperIII, SgrAPaperIV, SgrAPaperV, SgrAPaperVI, SgrAPaperVII, SgrAPaperVIII}. These images reveal a ring-like structure with an ordered linear polarization pattern, consistent with synchrotron emission from relativistic plasma accreting onto the black hole. Interpretation of these observations relies on general relativistic magnetohydrodynamics (GRMHD) simulations of black hole accretion \citep{narayan_jets_2022, dhruv_v3_grmhd_survey_2025}. Synthetic images and spectra produced from these simulations, when compared with EHT data and observations of \sgra at other wavelengths, constrain the state of the accreting plasma and the spacetime in the vicinity of the black hole \citep{SgrAPaperV,SgrAPaperVI,SgrAPaperVIII}. 

As part of the 2017 EHT campaign, the Atacama Large Millimeter/submillimeter Array (ALMA) recorded long-duration (3-10 hrs), high-cadence (4 s) 230 GHz light curves of the Galactic Center \citep{wielgus_mm_lightcurves_2022}. The measured source variability, characterized by the modulation index $\sigma/\mu$, was consistent with previous 230 GHz measurements of \sgra. However, most simulations in the EHT analysis exhibited a higher modulation index, meaning they were more variable than the actual source \citep{SgrAPaperV}. One possible explanation for this discrepancy, the \textit{variability crisis}, is missing physics in the numerical models.

In particular the simulations used to interpret EHT observations use an ideal GRMHD (IGRMHD) model, which treats the relativistic plasma as a fluid in local thermodynamic equilibrium \citep{gammie_harm_2003, mizuno_rezzolla_grmhd_chapter_2024}. Accretion flows surrounding low-luminosity active galactic nuclei (LLAGNs) like \sgra are, however, Coulomb-collisionless \citep{mahadevan_quataert_1997}, which opens the possibility that finite mean free path effects may alter both accretion dynamics and horizon-scale emission \citep{galishnikova_grpic_2023, galishnikova_anisotropic_images_2023}. For example, magnetic reconnection, a mechanism considered to explain high-energy flares observed in black hole accretion systems \citep{nathanail_plasmoid_grmhd_flares_2020, ripperda_flares_2022, hakobyan_radiative_reconnection_flares_2023, von_fellenburg_mir_flare_2025, solanki_reconnection_lightcruves_2025, sironi_relativistic_reconnection_2025}, is slower in the collisional regime compared to the collisionless expectation. The hierarchy of scales motivates a kinetic treatment of the problem, such as particle-in-cell (PIC) methods, that resolve the relevant microscopic scales in the plasma. Indeed, recent multidimensional, global, kinetic studies have explored black hole magnetospheres in the force-free limit \citep{parfrey_grpic_2019, crinquand_pair_discharges_2020, crinquand_gammaray_lightcurves_2021, crinquand_reconnection_radiation_2022, mellah_magnetosphere_disk_2022, mellah_3d_magnetosphere_2023} and have also modeled black hole accretion \citep{galishnikova_grpic_2023, vos_grpic_2025}. However, these simulations are too computationally expensive to permit the large-scale parameter surveys necessary for systematic comparison with observational data. Additionally, collective plasma phenomena such as pitch-angle scattering due to kinetic instabilities (see e.g., \citealp{bott_chapman_enskog_plasmas_2024} and references therein) and stochastic plasma echoes that stifle phase mixing in turbulent systems \citep{parker_sotchastic_echos_drift_kinetic_plasma_2016, schekochihin_stochastic_plasma_echos_2016, meyrand_fluidization_2019} tend to make the plasma more fluid-like.  In this work we model the plasma as a \textit{weakly collisional}, or equivalently, a \textit{dissipative} fluid, where non-ideal effects are introduced as deviations from thermodynamic equilibrium.

The theory of relativistic dissipative fluids is intricate, with some models prone to pathologies such as acausality and instability \citep{hiscock_lindblom_causality_1983,hiscock_lindblom_instabilities_1985,hiscock_lindblom_pathologies_1988,garcia_instabilities_2009,denicol_rischke_microscopic_foundations_2021}. Over the past few decades, some advancements were driven by the need to model quark-gluon plasma formed in relativistic heavy-ion collisions \citep{rocha_relativistic_dissipative_2024,shen_hydro_HIC_2020,romatschke_ten_years_relativistic_fluids_2017,bernhard_viscosity_QGP_2019,noronha_transport_coefficients_2009,heinz_viscosity_RHIC_2013,gale_hydro_HIC_2013}. The framework was also applied in cosmology, \citep{padmanabhan_chitre_viscous_1987, zakari_viscous_cosmo_1993, maartens_dissipative_cosmo_1995, piattella_bulk_viscous_2011, breivik_negative_viscosity_2013, breivik_relativistic_viscous_universe_2014, breivik_viscous_cosmology_2017}, and more recently, to the study of ultradense matter formed during neutron stars mergers (e.g., \citealp{most_micro_bulk_2024, most_anisotropy_eos_neutron_stars_2025}), and the dynamics of radiatively-inefficient accretion disks around supermassive black holes \citep{chandra_emhd_2015,foucart_2d_egrmhd_2016,foucart_3d_egrmhd_2017}.

In this Letter, we use the Extended GRMHD (EGRMHD) model of  \cite{chandra_emhd_2015} in global 3D simulations of black hole accretion. EGRMHD modifies ideal GRMHD by including heat conduction along magnetic field lines and shear viscosity. The theory is causal and strongly hyperbolic \citep{cordeiro_emhd_bounds_2024}. This Letter  goes beyond \cite{foucart_3d_egrmhd_2017} by generating synthetic horizon-scale images and spectral energy distributions (SEDs) from a new set of high-resolution, long-duration EGRMHD simulations. By comparing images and SEDs with those from corresponding IGRMHD simulations, we evaluate the impact of dissipative physics on electromagnetic observables. We find that time-averaged images and spectra are almost unchanged, although simulations incorporating dissipative effects produce 230 GHz light curves with reduced variability on three-hour timescales. 

\section{Methods}
\label{sec:methods}

We have modified the ideal, GPU-enabled GRMHD code \kharma \citep{prather_kharma_2024} to simulate weakly collisional accretion onto a black hole. The code evolves two additional scalar variables: the scalar heat flux along the magnetic field, $q$, and the pressure anisotropy defined with respect to the local magnetic field $\Delta P$ ($\Delta P=P_{\perp} - P_{\parallel}$). These are defined as $q^{\mu}\equiv q\hat{b}^{\mu}$ and $\pi^{\mu\nu}\equiv-\Delta P(\hat{b}^{\mu}\hat{b}^{\nu}-\frac{1}{3}h^{\mu\nu})$, where $q^{\mu}$ is the heat flux four-vector, $\hat{b}^{\mu}$ is the unit  magnetic field four-vector, $\pi^{\mu\nu}$ is the shear-stress tensor, and $h^{\mu\nu}$ is the projection tensor onto a spatial slice orthogonal to the fluid four-velocity $u^{\mu}$. The evolution equations for $q$ and $\Delta P$ include source terms with time derivatives, requiring a locally semi-implicit time-stepping scheme (\citealt{chandra_grim_2017}; see Appendix \ref{appendix:egrmhd_in_kharma} for details on the algorithm used in EGRMHD simulations and a suite of test problems validating the implementation in \kharma).  EGRMHD simulations are $\sim10$x more expensive than their ideal counterparts.

\begin{figure*}
\centering
\includegraphics[,width=\linewidth]{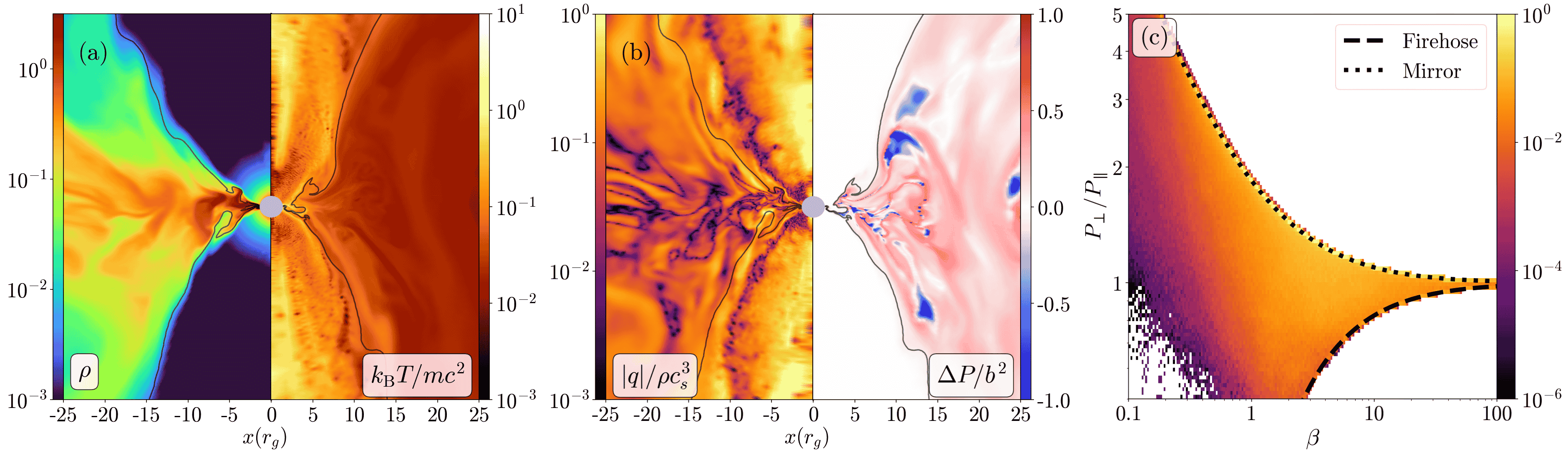}
\caption{Snapshot from an EGRMHD, MAD, $a_{*} = 15/16$ simulation. (a) Poloidal $(r,\theta)$ slices of fluid rest-mass density (left) and dimensionless temperature (right). (b) Poloidal slices of heat flux normalized by the free-streaming value (left) and pressure anisotropy normalized by magnetic energy density (right). The black contour marks $\sigma = 1$, which separates the accretion disk from the magnetically dominated jet. (c) Mass-weighted distribution of pressure anisotropy as a function of $\beta$ within $r \leq 20r_{\text{g}}$ for the snapshot shown in panels (a) and (b). The black dotted (dashed) lines indicate the mirror (firehose) instability threshold.}\label{fig:emhd_snapshot}
\end{figure*}

Our simulations are initialized with a hydrostatic equilibrium torus solution \citep{fishbone_relativistic_1976}. The solution has two free parameters: the radius at the inner edge of the disk $r_{\text{in}}$ and the pressure maximum radius $r_{\text{max}}$. We seed the torus with a poloidal magnetic field, and as the fluid accretes, the magnetic field is dragged along (consistent with Alfv\'{e}n's theorem, \citealp{alfven_EM_hydro_waves_1942}), causing magnetic flux to accumulate on the event horizon. The accumulated flux is characterized by the dimensionless flux $\phi_b\equiv\Phi_{\text{BH}}/(\dot{M}r_g^2c)^{1/2}$ (here $\Phi_{\text{BH}}$ is the net magnetic flux crossing one hemisphere of the event horizon; \citealp{tchekhovskoy_efficient_2011}).

The evolution exhibits two distinct states, depending on $\phi_b$. In the magnetically-arrested state (MAD; \citealp{narayan_mad_2003, igumenshchev_2003, tchekhovskoy_efficient_2011}) $\phi_b \sim 16$ and magnetic flux grows until it is large enough to halt accretion.  Flux is then expelled in a violent eruption event, and flux accumulation begins again. In the standard and normal evolution state (SANE; \citealp{narayan_sane_2012, sadowski_sane_2013}) $\phi_b\ll16$ and the magnetic field is comparatively weak and drives outward transport of angular momentum via the magnetorotational instability (MRI). Our initial magnetic field is expressed in terms of a vector potential $A_{\phi}$, which is $\text{max}[\rho/\rho_{\text{max}}(r/r_{\text{in}}\,\text{sin}\theta)^3 e^{-r/400}-0.2,\,0]$ for MAD models and $\text{max}[\rho/\rho_{\text{max}}-0.2,\,0]$ for SANE models. Here, $\rho$ is the fluid rest-mass density and $\rho_{\text{max}}$ is the maximum density in the initial torus.

We consider four EGRMHD simulations: SANE and MAD at black hole spin  $a_{*}=0,15/16$ ($a_{*}\equiv Jc/(GM^2)$; hereafter we adopt units such that $GM=c=1$). We also conduct four otherwise identical IGRMHD simulations as controls. The governing equations are solved in  modified spherical Kerr-Schild coordinates (FMKS; \citealp{ Wong_2022_patoka}) which concentrates grid zones at the midplane close to the event horizon. The computational domain has $N_r \times N_\theta \times N_\phi = 384 \times 192 \times 192$ resolution elements.  The grid extends radially from just inside the event horizon to $1000\,r_{\text{g}}$, with $\theta \in [0, \pi]$ and $\phi \in [0, 2\pi]$.  Here $r_{\text{g}} \equiv GM/c^2$ is the gravitational radius. Each simulation is evolved to $t=30,000\,t_{\text{g}}$ ($t_{\text{g}}\equiv~r_{\text{g}}/c$), which is long enough to enable a comparison of model and observed light curve variability  \citep{lee_gammie_windowed_gaussian_2022,SgrAPaperV,wielgus_mm_lightcurves_2022}.

Our GRMHD simulations evolve a single fluid, with electron temperature determined from a parameterized model in the radiative transfer calculations.  We assume a thermal electron distribution function and  prescribe the electron temperature using the ``$R_{\text{high}}$'' model \citep{moscibrodzka_rhigh_2016}, which is motivated by models of kinetic dissipation of Alfv\'{e}nic turbulence \citep{quataert_alfvenic_heating_1998, quataert_gruzinov_particle_heating_1999, howes_prescription_2010,gammie_adiabatic_index_2025}. To generate synthetic images we use the polarized ray-tracing code \ipole \citep{moscibrodzka_ipole_2018}. To generate SEDs we use the Monte Carlo radiation transport code \igrmonty \citep{dolence_igrmonty_2009} which accounts for synchrotron, bremsstrahlung, and Compton scattering. The simulation suite is summarized in Table \ref{table:simulation_parameters}.

\setlength{\tabcolsep}{5pt}
\begin{table}[ht]
\centering
\caption{Summary of GRMHD and GRRT parameters for EGRMHD and IGRMHD simulations. $\hat{\gamma}$ is the adiabatic index of the fluid \footnote{The adiabatic index was chosen to enable comparison with earlier EHT-related simulations \citep{Wong_2022_patoka, dhruv_v3_grmhd_survey_2025}; a better choice for two-temperature, collisionless accretion flows would be $\hat{\gamma}$ slightly less than 5/3 \citep{gammie_adiabatic_index_2025, chael_two_temp_radgrmhd_survey_2025}.}; $r_{\text{in}}$ ($r_{\text{max}}$) is the inner (pressure-maximum) radius of the initial torus. $R_{\text{high}}$ is a free parameter in the emission model that sets the ion-to-electron temperature ratio;  $i$ is the inclination angle (angle between the line of sight and the black hole spin axis). All synthetic images have a $200\mu as$ field of view.} \label{table:simulation_parameters}
\begin{tabular}{ccccccc}
\hline\hline
Flux & $a_*$ & $\hat{\gamma}$ & $r_{\text{in}}$ & $r_{\text{max}}$ & $R_{\text{high}}$ & $i$($\degree$) \\
\hline
MAD & 0 & 13/9 & 20 & 41 & 1,10,40,160 & 10,30,...,90 \\
MAD & 15/16 & 13/9 & 20 & 41 & 1,10,40,160 & 10,30,...,90 \\
SANE & 0 & 4/3 & 10 & 20 & 1,10,40,160 & 10,30,...,90 \\
SANE & 15/16 & 4/3 & 10 & 20 & 1,10,40,160 & 10,30,...,90 \\
\hline
\end{tabular}
\end{table}

\section{Results}
\label{sec:results}

The initial state evolves due to (1) winding of initially radial field lines by differential rotation and (2) the MRI \citep{MRI_I_1991, MRI_II_1991, MRI_III_1992, MRI_IV_1992, HGB_MRI_1995, balbus_hawley_review_1998}. SANE models do not accumulate significant magnetic flux on the horizon and flow dynamics are governed primarily by fluid forces, with $\beta \sim 10$ in the disk ($\beta\equiv P_{\text{gas}}/P_{\text{mag}}$ is the ratio of fluid pressure to magnetic pressure). MAD models accumulate significan magnetic flux, generating strong, ordered magnetic fields ($\beta\sim1$ in the disk near the black hole), relativistic jets along the spin axis, and  intermittent flux-eruption events \citep{tchekhovskoy_efficient_2011,ripperda_flares_2022,chatterjee_angular_momentum_2022,gelles_flux_eruption_2022}.  MAD models are favored for EHT sources \citep{M87PaperV, SgrAPaperV}. These features of SANE and MAD models are also observed in our EGRMHD simulations.

The non-ideal fields are initialized to zero but evolve on the dynamical time $\tau_{d}\equiv (r^{3}/GM)^{1/2}$ toward their corresponding (covariant) Braginskii values \citep{braginskii_transport_1965}. The pressure anisotropy cannot grow unbounded and is limited by the onset of mirror \citep{hasegawa_mirror_1969,southwood_kivelson_mirror_1993,kivelson_southwood_mirror_1996,Kunz_2014} and firehose \citep{rosenbluth_firehose_1957,chandrasekhar_firehose_1958,parker_anisotropic_gas_1958,gary_proton_firehose_1998,bott_firehose_highbeta_2025} instabilities, which pitch-angle scatter the particles. The EGRMHD model incorporates this effect by increasing the effective scattering rate as the instability boundaries are approached.  The effect is to confine $\Delta P$ within the mirror and firehose bounds, as suggested by PIC simulations \citep{Kunz_2014}. The model incorporates a similar increase in scattering rate as the heat flux approaches the free-streaming value $q_{\text{max}}\simeq\rho c_s^{3}$ ($c_s$ is the sound speed).



\begin{figure}
\centering
\includegraphics[,width=\linewidth]{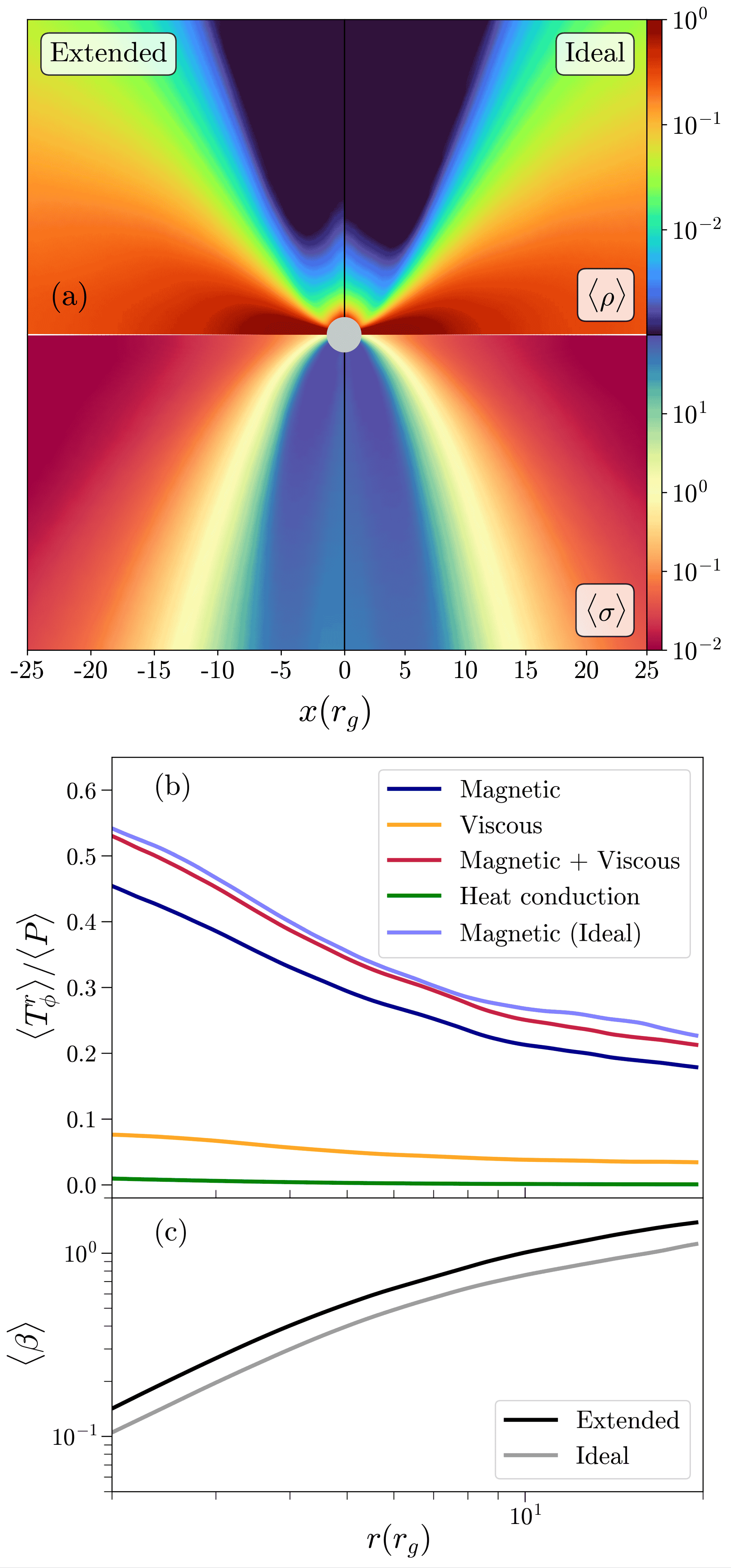}
\caption{Comparison of time-averaged fluid quantities between Extended and Ideal simulations (MAD, $a_{*} = 15/16$). (a) The top row shows the time- and azimuthally-averaged rest-mass density, $\rho$, and the bottom row displays plasma magnetization, $\sigma$. The left column shows the Extended simulation and the right column shows the Ideal simulation. (b) Angular momentum transport in the disk: components of the density-weighted average $\langle T^{r}_{\phi}\rangle$ normalized by gas pressure $\langle P\rangle$ are plotted as a function of radius. (c) Radial profiles of $\beta$ for Extended and Ideal simulations.}\label{fig:emhd_imhd_comparison}
\end{figure}

\begin{figure*}
\centering
\includegraphics[,width=\linewidth]{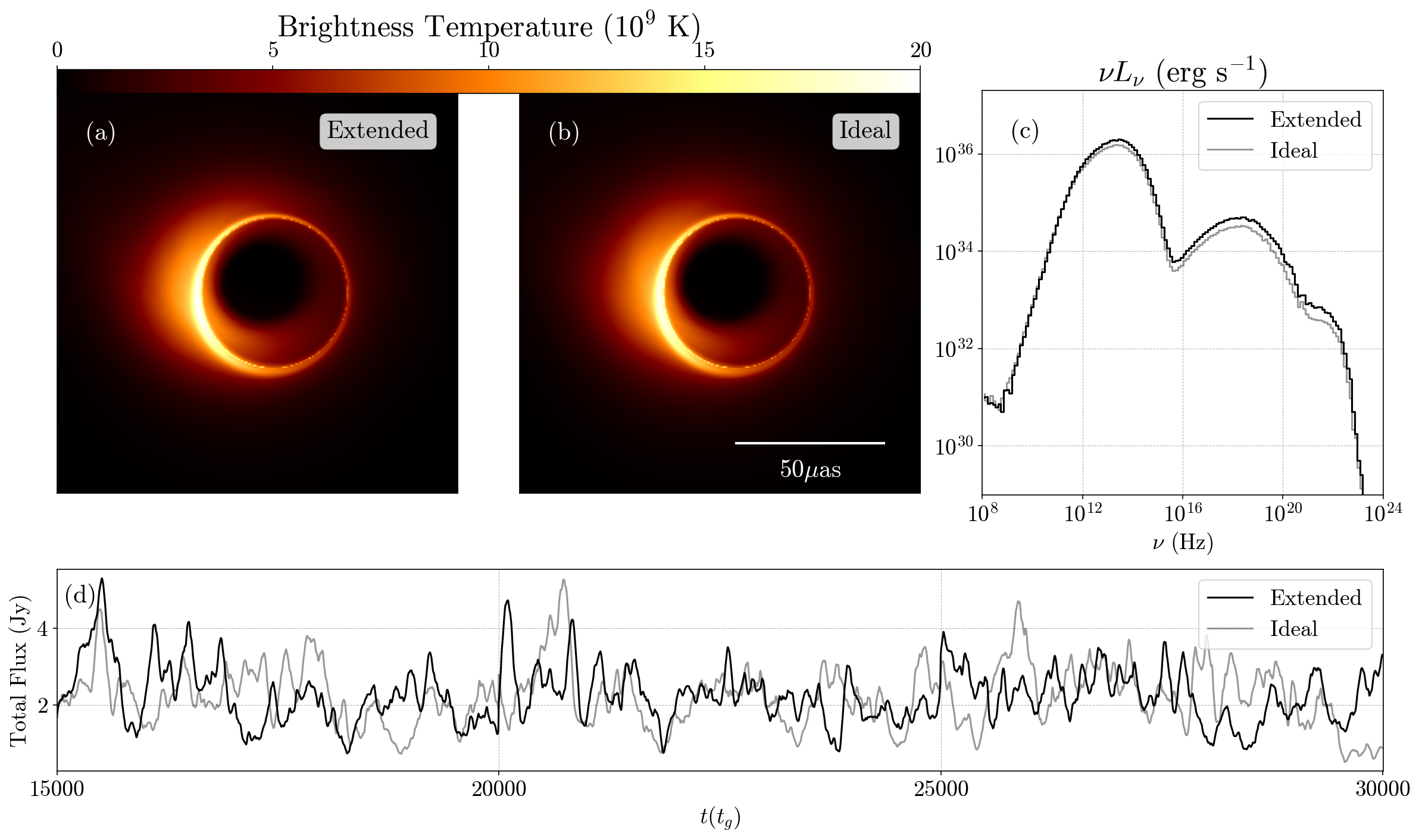}
\caption{Electromagnetic observables for MAD $a_{*}=15/16$ simulation with $R_{\text{high}}=160$ at a viewing angle of $30\degree$. (a) and (b) Time-averaged 230 GHz total intensity images for the Extended and Ideal GRMHD simulations, respectively. (c) Comparison of time-averaged spectra for the simulations shown in (a) and (b). (d) Light curves at 230 GHz over a duration of $\Delta t = 15,000~t_{\text{g}}$ ($\sim84$ hours for \sgra).}
\label{fig:grrt_comparison}
\end{figure*}

Figure \ref{fig:emhd_snapshot} shows a snapshot of the MAD $a_{*}=15/16$ simulation at $t\sim30,000\,t_{\text{g}}$. At this point a quasi-steady state is well established in the inner regions of the accretion flow. We find that the heat flux remains well below its free-streaming value near the disk midplane, with $q\lesssim0.1\,q_{\text{max}}$, and has a negligible impact on the flow's thermodynamics. An appreciable fraction of the disk mass lies near near the mirror and firehose thresholds: $\Delta P>0.99\Delta P_{\text{mirror}}$ or $\Delta P>0.99\Delta P_{\text{firehose}}$, where $\Delta P_{\text{mirror}}\equiv b^{2}/2\cdot(P_{\parallel}/P_{\perp})$ ($\sim b^{2}/2$ when $\beta\gg1)$, and $\Delta P_{\text{firehose}}\equiv-b^{2}$. Within the inner $20\,r_{g}$, 40-45\% of the plasma reaches the mirror threshold and 2-3\% is at the firehose threshold. Panel (c) shows the mass-weighted distribution of pressure anisotropy in the $(\beta, P_{\perp}/P_{\parallel})$ plane. Notably, this tendency for saturation at the instability thresholds is more pronounced in SANE simulations where $\sim65\%$ of the disk resides at mirror threshold and $\sim10\%$ at the firehose threshold (see Appendix \ref{appendix:sane_vs_mad_egrmhd} for a more detailed comparison of SANE and MAD simulations).

We model mirror and firehose instabilities as mechanisms that regulate the growth of pressure anisotropy, motivated by solar wind measurements \citep{hellinger_solar_wind_anisotropy_2006, bale_solar_wind_anisotropy_2009} which show that for $\Delta P > 0$ the plasma anisotropy is bound by the the mirror instability threshold, and exceeds the predicted ion cyclotron (IC) threshold. However, this could be due to the assumption of a bi-Maxwellian plasma when calculating the IC threshold \citep{isenberg_icw_2012, isenberg_icw_threshold_2013}. PIC simulations of magnetically-dominated ($\beta\lesssim1$) electron-ion plasmas that are motivated by MAD accretion flows find that the anisotropy of each species in the saturated state is predominantly set by its respective cyclotron instability (Dhruv et al., in preparation). In a future study we will incorporate the IC threshold in our equations to study its potential importance on the flow dynamics.



Figure \ref{fig:emhd_imhd_comparison} compares the time-averaged structures of the MAD $a_{*}=15/16$ IGRMHD and EGRMHD simulations. The models are remarkably similar, as shown in panel (a), which shows the azimuthally averaged profiles of $\rho$ and plasma magnetization $\sigma \equiv b^{2}/\rho$ averaged over $15,000$ to $30,000\,t_{\text{g}}$. In MAD models strong magnetic fields govern the dynamics, suppressing the influence of viscous stresses and heat conduction. Although MAD models exhibit  larger pressure anisotropy $\Delta P/P$ than SANE models (see Figure \ref{fig:emhd_qdP_distribution}), the anisotropy remains small, on average, relative to magnetic pressure. The pressure anisotropy is equivalent to a viscosity, and its inclusion provides an additional mechanism for angular momentum transport through the shear stress $\sim-\Delta P\,\hat{b}^{r}\hat{b}_{\phi}$. In EGRMHD simulations the combined magnetic and viscous angular momentum flux is comparable to the magnetic flux in IGRMHD simulations (see panel (b) in Figure \ref{fig:emhd_imhd_comparison}) and the EGRMHD disk is $20-30\%$ less magnetized (see panel (c) in Figure \ref{fig:emhd_imhd_comparison}). In summary, we find that the time-averaged structure of the accretion flow is remarkably similar in weakly collisional and ideal models, despite the rapid growth of pressure anisotropy.

\begin{figure}
\centering
\includegraphics[,width=\linewidth]{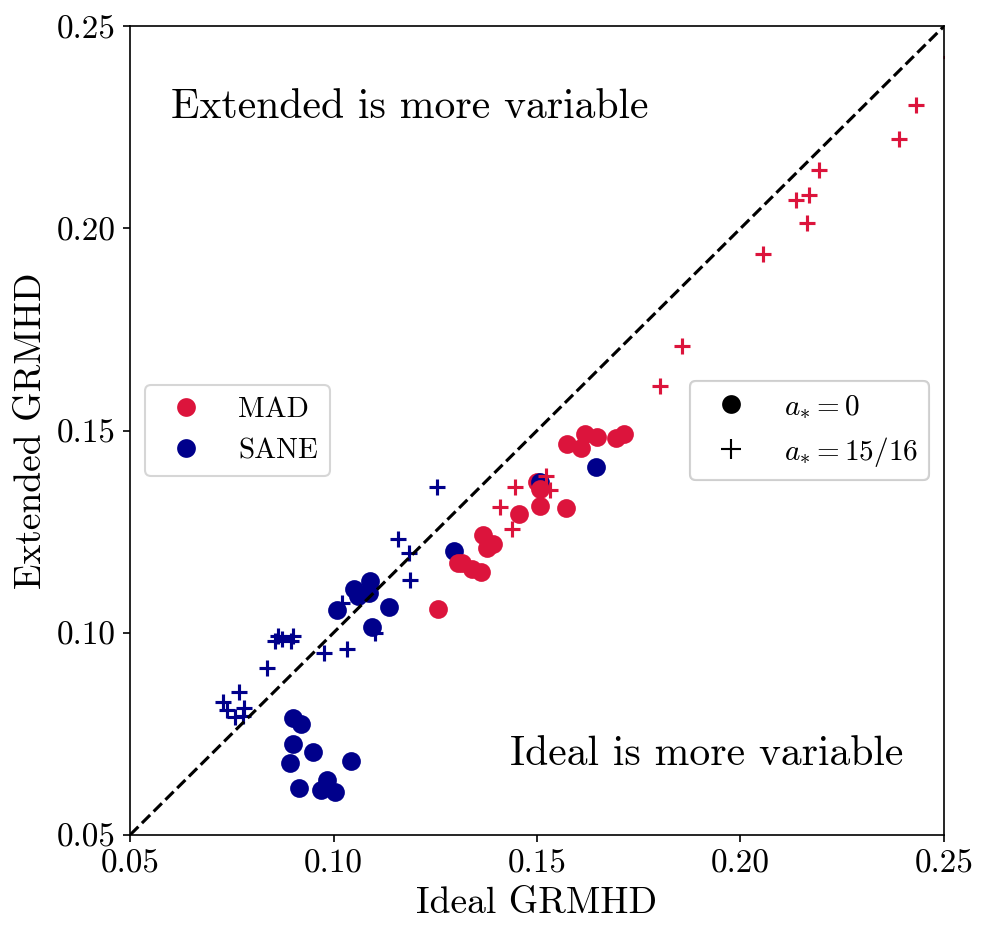}
\caption{Three-hour modulation index for all models considered in this work. Marker color represents the magnetization state (MAD vs SANE), while marker shape indicates the black hole spin. Points below the dashed line represent models where the Extended GRMHD simulation produces 230 GHz light curves with lower variability than their Ideal counterparts.}\label{fig:mi_comparison}
\end{figure}


Figure \ref{fig:grrt_comparison} shows electromagnetic observables for one of EHT's preferred models of the Galactic center \citep{SgrAPaperV}---MAD, $a_{*}=15/16$, $R_{\text{high}}=160$, and an inclination angle of $30\degree$. Panels (a) and (b) plot total intensity images for the EGRMHD and IGRMHD simulations, respectively, averaged over $5000\,t_{\text{g}}$, while panel (c) compares their SEDs. The time-averaged radiative signatures of the weakly collisional fluid models are nearly indistinguishable from those of the ideal models. We attribute this to the similarity in fluid structures between the two plasma models, which governs the ion-to-electron temperature ratio $T_{i}/T_{e}$ for the chosen emission model, along with the assumption of an isotropic Maxwellian electron distribution function. This strong similarity in synthetic observables is a general feature of our models.

Submillimeter-wavelength observations of \sgra suggest that its light curve can be modeled as a red noise process on timescales ranging from minutes to a few hours \citep{dexter_submm_variability_2014, georgiev_variability_2022, wielgus_mm_lightcurves_2022}. EHT analyses of \sgra \citep{SgrAPaperV} characterized the light curve variability using the three-hour modulation index $M_{3}$ (where $M_{\Delta t}\equiv\sigma_{\Delta t}/\mu_{\Delta t}$, $\sigma_{\Delta t}$ is the standard deviation measured over the interval $\Delta t$, and $\mu_{\Delta t}$ is the mean over the same interval). All ideal GRMHD MAD models, along with a significant fraction of SANE models, were found to exhibit excess variability compared to observations. 

Figure \ref{fig:mi_comparison} highlights the variability trends in EGRMHD and IGRMHD simulations. We analyzed independent three-hour segments of the light curve over $15,000\,t_{\text{g}}$ (from $t=15,000$ to $30,000 \,t_{\text{g}}$; see panel (d) in Figure \ref{fig:grrt_comparison}), corresponding to $\sim 83$ hours for \sgra and yielding a sample size of 27 intervals for the $M_{3}$ analysis. We find that $\approx75\%$ of weakly collisional models (marginalized over all GRMHD and GRRT parameters) exhibit lower variability than their ideal counterparts. Although this reduction  does not fully reconcile the discrepancy with observations (e.g., \citealp{wielgus_mm_lightcurves_2022} report $M_{3}\in[0.024,0.051]$ from April 5-10, 2017), it is notable that \textit{all} MAD EGRMHD simulations show a systematic decrease in variability. The power spectral density (PSD) of the lightcurves indicates that both EGRMHD and IGRMHD simulations exhibit similar slopes at timescales $\lesssim 2$ hours. EGRMHD simulations with lower $M_{3}$ relative to the ideal case generally show reduced variability at timescales longer than $\sim 1$ hour. 

Although variations in electron anisotropy might change light curve variability that is not possible here because we have assumed an isotropic electron distribution.  The reduced variability of EGRMHD models must therefore be caused by changes in the EGRMHD fluid evolution.  This is plausibly explained by lower turbulent intensity in EGRMHD models, as evidenced by higher average $\beta$ (see Figure \ref{fig:emhd_imhd_comparison}): weaker field implies weaker velocity and density fluctuations, lowering light curve variability.


\section{Summary}
\label{sec:summary}

We have studied the impact of low collisionality on the structure and observables associated with low luminosity black hole accretion. We used the Extended GRMHD model \citep{chandra_emhd_2015}, which incorporates leading order corrections to the ideal GRMHD model: pressure anisotropy, or equivalently viscosity, and heat conduction. We consider both MAD and SANE (strongly and weakly magnetized) models. We find the flow structure in the Extended and Ideal GRMHD models to be similar. The only significant difference we observe is a higher level of magnetization in the IGRMHD disk. These results are consistent with previous work \citep{foucart_3d_egrmhd_2017}, which integrated the same physical model at lower numerical resolution (up to 6x). Whether these findings persist at even higher numerical resolutions is an open question, as viscous shear stresses do seem to be sensitive to resolution in a local, nonrelativistic Braginskii model \citep{kempski_shearing_box_2019}.

We have also presented the first event-horizon–scale images and spectra of weakly collisional accretion models of \sgra. We find 230 GHz lightcurves of \sgra from weakly collisional, magnetically dominated models to be less variable than their ideal counterparts, resulting in better agreement with observations. The time-averaged properties of the models' radiative signatures, however, strongly resemble corresponding IGRMHD simulations. 

We attribute the similarity of weakly collisional and ideal models to (i) pressure-anisotropy limiters that model the effect of plasma instabilities, thereby limiting the influence of pressure anisotropy; and (ii) the use of a simple, isotropic, thermal electron distribution function in estimating the emergent radiation. For example, \cite{salas_variability_2025} showed that radiative cooling in two-temperature models reduces light-curve variability. In addition, in magnetized, collisionless plasmas, electrons naturally develop anisotropies which are regulated by nonresonant instabilities such as mirror and firehose, and by resonant kinetic instabilities, e.g., whistler \citep{sudan_whistler_1963, sudan_whistler_1965, gladd_relativistic_whistler_1983}. Anisotropic eDFs can directly affect horizon-scale synthetic observables \citep{galishnikova_anisotropic_images_2023} and viscous stresses may be a dominant source for electron heating in collisionless disks \citep{sharma_electron_heating_2007}. Moreover, nonthermal processes in the accretion flow can generate power-law tails (see e.g., \citealp{comisso_pitch_angle_turbulence_2021, comisso_ion_electron_acceleration_turbulence_2022, comisso_pitch_angle_reconnection_2024}), which---together with electron anisotropy---may help explain limb-brightened jet images of M87* \citep{tsunetoe_limb_brightening_2025}.

Looking forward, it would be natural to extend the EGRMHD framework to a full two-fluid model that solves separate evolution equations for electrons and ions. The model would self-consistently predict electron energy density, heat flux, and pressure anisotropy.  Any such model is likely to be very expensive, however, and if they are to be truly predictive then the closure relations (e.g. estimates for the viscosity and heat conductivity) would have to be calibrated by kinetic (PIC) simulations.


\begin{acknowledgements}
V.D. is grateful to Eliot Quataert, Alisa Galishnikova, and Jixun Ding for helpful discussions, and to Harshi Rambhia for invaluable assistance with the figures. We also thank an anonymous referee for comments that improved the manuscript. V.D. was supported in part by a Dissertation Completion Fellowship, a Donald C.
and F. Shirley Jones Fellowship and an  ICASU/NCSA Fellowship.
This work was supported by NSF grant AST 20-34306. This research is part of the Delta research computing project, which is supported by the National Science Foundation (award OCI 2005572), and the State of Illinois. Delta is a joint effort of the University of Illinois at Urbana-Champaign and its National Center for Supercomputing Applications. This research used resources of the Oak Ridge Leadership Computing Facility at the Oak Ridge National Laboratory, which is supported by the Office of Science of the U.S. Department of Energy under Contract No. DE-AC05-00OR22725. The data analysis was possible thanks to the high throughput computing utility `Launcher' \citep{wilson_launcher_2017}.
\end{acknowledgements}

\appendix

\section{Extended GRMHD in \kharma}
\label{appendix:egrmhd_in_kharma}

\begin{figure}
\centering
\includegraphics[,width=0.9\linewidth]{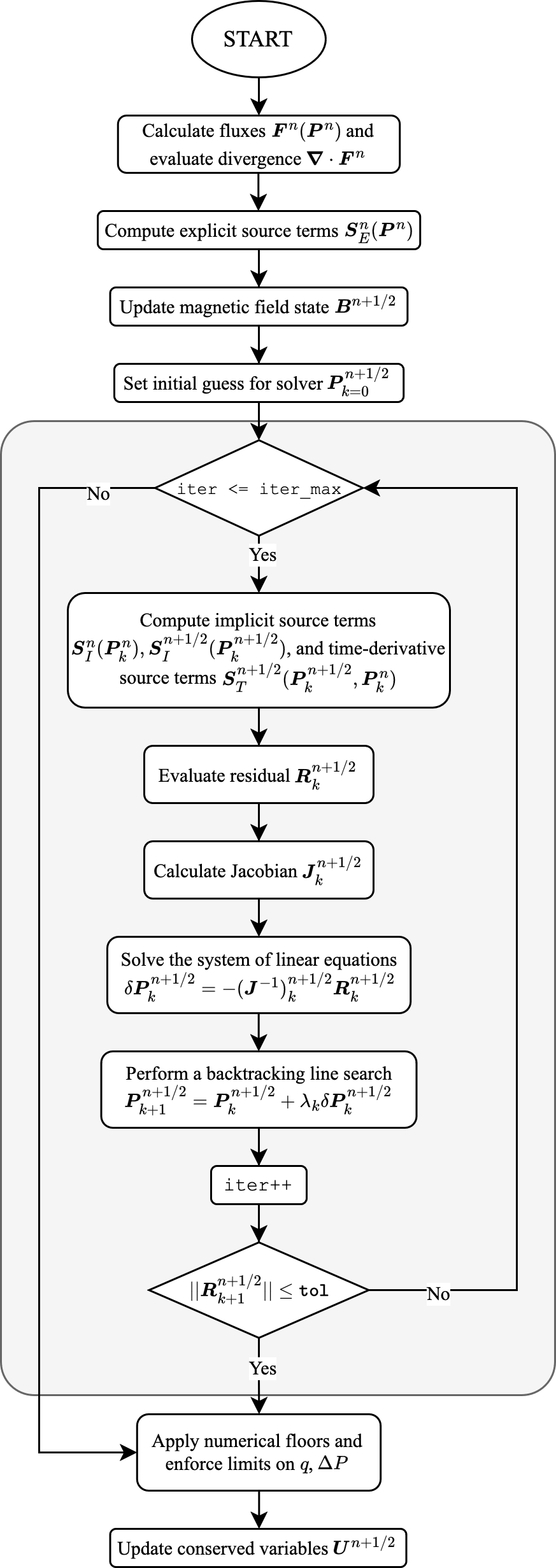}
\caption{A schematic flowchart illustrating the sequence of operations during a half-step ($t^{n}\rightarrow t^{n+1/2}$) in the EGRMHD evolution. The gray box marks the \texttt{Implicit} kernel, which iteratively determines the next fluid state.}
\label{fig:emhd_flowchart_kharma}
\end{figure}

\begin{figure*}
\centering
\includegraphics[,width=\linewidth]{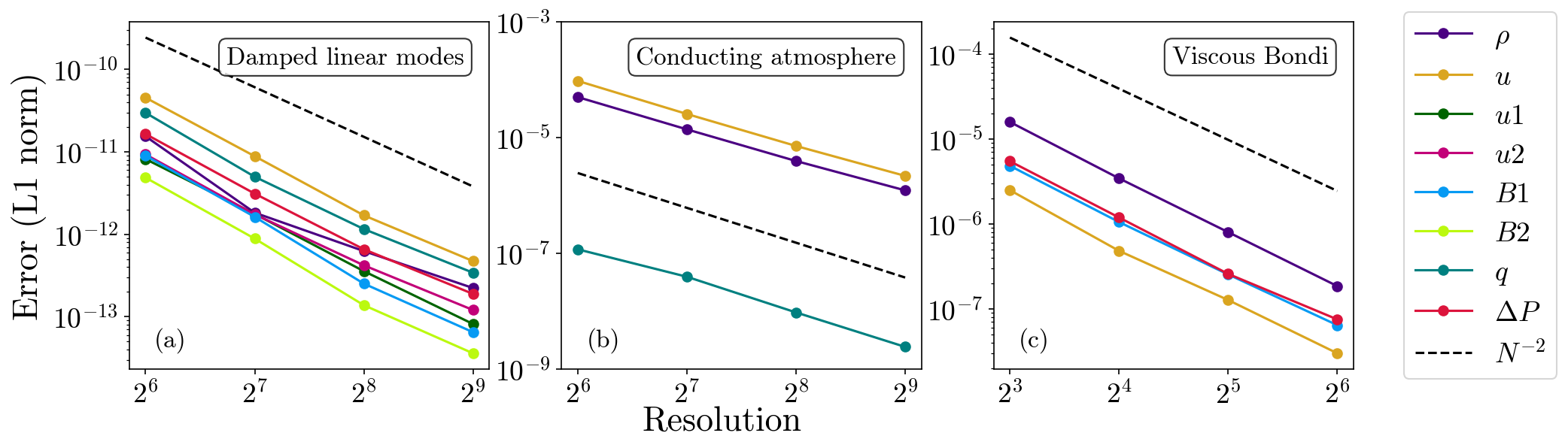}
\caption{Convergence tests for EGRMHD test problems. (a) Damped linear waves in flat space. (b) Hydrostatic equilibrium in Schwarzschild geometry with a radial temperature gradient. (c) Spherical accretion with anisotropic viscosity, neglecting the backreaction of viscosity on the steady-state inflow.}
\label{fig:emhd_convergence_kharma}
\end{figure*}

In this section we briefly discuss the EGRMHD model, summarize its implementation in \kharma, and present results from a suite of test problems that validate its numerical implementation.

\subsection{Physical Model}

The EGRMHD formalism is a single-fluid description of plasmas that satisfy the hierarchy of length scales $\rho_{\text{L}}\ll\lambda_{\text{mfp}}\ll r_{\text{g}}$, where $\rho_{\text{L}}$ is the particle Larmor radius and $\lambda_{\text{mfp}}$ is the collisional mean free path. This regime implies a collisional plasma with anisotropic transport along the local magnetic field. EGRMHD is a relativistic generalization of the Braginskii model \citep{braginskii_transport_1965}, replacing constitutive expressions for heat flux and pressure anisotropy with evolution equations,\begin{subequations}\label{eqn:emhd_original_equations}
\begin{align}
    \frac{dq}{d\tau} &= -\frac{q - q_{0}}{\tau_{R}} - \frac{q}{2}\frac{d}{d\tau}\text{log}\bigg(\frac{\tau_{R}}{\chi P^{2}}\bigg) \label{eqn:emhd_original_equations_q}\\
    \frac{d\Delta P}{d\tau} &= -\frac{\Delta P - \Delta P_{0}}{\tau_{R}} - \frac{\Delta P}{2}\frac{d}{d\tau}\text{log}\bigg(\frac{\tau_{R}}{\rho\nu P}\bigg). \label{eqn:emhd_original_equations_dP}
\end{align}
\end{subequations}Here $\chi,\,\nu$ are thermal and momentum diffusivity respectively, and $q_{0}$ and $\Delta P_{0}$ are the covariant generalizations of the Braginskii heat flux and pressure anisotropy $q_{0}\equiv-\rho\chi\hat{b}^{\mu}(\nabla_{\mu}\Theta + \Theta a_{\mu})$ and $\Delta P_{0}\equiv3\rho\nu(\hat{b}^{\mu}\hat{b}^{\nu}\nabla_{\mu}u_{\nu} - 1/3\cdot\nabla_{\mu}u^{\mu})$. $\tau_{R}$ is a relaxation timescale that dictates how quickly $q$ and $\Delta P$ approach their respective Braginskii values. In the collisionless regime, $\tau_{R}$ may be interpreted as the effective mean free time due to wave-particle scattering. $\tau$ denotes the proper time, and the operator $d/d\tau\equiv u^{\mu}\nabla_{\mu}$ is the relativistic extension of the material (convective) derivative. The evolution   equations for $q$ and $\Delta P$ are derived from thermodynamic considerations using an Israel-Stewart-like approach \citep{israel_stewart_1979}, and are closed by constitutive relations for $\chi$ and $\nu$ from nonrelativistic collisional theory: $\chi=\phi c_{s}^{2}\tau_{\text{R}}$, $\nu=\psi c_{s}^{2}\tau_{\text{R}}$. Here, $c_s$ is the relativistic sound speed, while $\phi$ and $\psi$ are dimensionless constants of order unity. These parameters govern the influence of the dissipative terms and are chosen to ensure the model remains causal and stable. The EGRMHD stress-energy tensor is given by,
\begin{equation}
    T^{\mu\nu} = T^{\mu\nu}_{\text{ideal}} + q^{\mu}u^{\nu} + q^{\nu}u^{\mu} + \pi^{\mu\nu},
\end{equation}
where $T^{\mu\nu}_{\text{ideal}}$ is the IGRMHD stress-energy tensor. A detailed description of the EGRMHD model is provided in \cite{chandra_emhd_2015}.

\subsection{Numerical Implementation and Validation}

\kharma \citep{prather_kharma_2024} is an open-source C++17 rewrite of the \harm algorithm \citep{gammie_harm_2003} designed to run efficiently on heterogeneous architectures. Originally designed for IGRMHD simulations of black hole accretion, the code fosters extensibility through a package-based framework that simplifies the addition of new physics. Below, we describe our approach to incorporating EGRMHD in \kharma. We adopt the \grim algorithm as described in \cite{chandra_grim_2017}.

In addition to the eight evolution equations corresponding to the ideal MHD framework that evolve $\rho$, $u^{\mu}$ and $\boldsymbol{B}$, EGRMHD evolves $q$ and $\Delta P$. For numerical stability, the code evolves Equations \ref{eqn:emhd_original_equations_q}, \ref{eqn:emhd_original_equations_dP} rescaled by $\rho$,
\begin{subequations} \label{eqn:emhd_rescaled_equations}
\begin{align}
    \nabla_{\mu}(\tilde{q}u^{\mu}) &= -\frac{\tilde{q} - \tilde{q}_{0}}{\tau_{R}} + \frac{\tilde{q}}{2}\nabla_{\mu}u^{\mu} \label{eqn:emhd_rescaled_equations_q}\\
    \nabla_{\mu}(\Delta\tilde{P}u^{\mu}) &= -\frac{\Delta\tilde{P} - \Delta\tilde{P}_{0}}{\tau_{R}} + \frac{\Delta\tilde{P}}{2}\nabla_{\mu}u^{\mu}, \label{eqn:emhd_rescaled_equations_dP}
\end{align}
\end{subequations}where $\tilde{q}=q(\tau_{R}/\rho\chi\Theta^{2})^{1/2}$ and $\Delta\tilde{P}=\Delta P(\tau_{R}/\rho\nu\Theta)^{1/2}$. Equations \ref{eqn:emhd_rescaled_equations_q}, \ref{eqn:emhd_rescaled_equations_dP} contain (i) stiff source terms, e.g., $\tilde{q}/\tau_{R}$ where $\tau_{R}$ can attain a very small value, and (ii) source terms with time derivatives such as $\tilde{q}_{0}\sim\nabla_{\mu}\Theta$, necessitating a local semi-implicit solver. The explicit timestepping scheme for IGRMHD is replaced by a semi-implicit scheme where the fluid variables are updated via a seven-dimensional Newton-Raphson solve, while the magnetic field is updated explicitly. 

Figure \ref{fig:emhd_flowchart_kharma} depicts the algorithm during half-step $t^{n}\rightarrow t^{n+1/2}$. The initial sequence of operations---calculating the face-centered fluxes $\boldsymbol{F}^{n}(\boldsymbol{P}^n)$ and their divergence $\boldsymbol{\nabla}\cdot\boldsymbol{F}^{n}$, evaluating explicit source terms $\boldsymbol{S}_{E}^{n}(\boldsymbol{P}^n)$, and updating the magnetic field primitives $\boldsymbol{B}^{n+1/2}$---are identical to an explicit update. $\boldsymbol{P}^n$ represents the vector of primitive variables at timestep `$n$'. The gray box indicates the series of tasks within the \texttt{Implicit} kernel that solves a system of nonlinear equations for the fluid primitives. The initial guess for the solver ($k=0$) is $\boldsymbol{P}^n$ along with the updated magnetic field primitives $\boldsymbol{B}^{n+1/2}$. The solver iteratively refines the estimate for $\boldsymbol{P}^{n+1/2}$ by finding the roots of the residual $\boldsymbol{R}^{n+1/2}\equiv(\boldsymbol{U}^{n+1/2}-\boldsymbol{U}^{n})/(\Delta t/2) + \boldsymbol{\nabla}\cdot\boldsymbol{F}^{n}-\boldsymbol{S}^{n}$ where $\boldsymbol{U}$ is the vector of conserved variables and $\boldsymbol{S}$ is the source term vector (it includes explicit $\boldsymbol{S}_{E}$, implicit $\boldsymbol{S}_{I}$ source terms and source terms that contain a time-derivative $\boldsymbol{S}_{T}$). This procedure is equivalent to solving the evolution equations. A backtracking line search is employed to ensure each iteration improves $\boldsymbol{P}^{n+1/2}$. Once the prescribed tolerance (\texttt{tol}) is reached or the solver exceeds the maximum iteration count (\texttt{iter\_max}), the code exits the \texttt{Implicit} kernel. $q$ and $\Delta P$ are adjusted to maintain $q<q_{\text{max}}$ and $\Delta P_{\text{firehose}}\leq\Delta P\leq\Delta P_{\text{mirror}}$. Finally, the half-step conserved variables $\boldsymbol{U}^{n+1/2}$ are computed. \kharma then advances the solution from $t^{n}\rightarrow t^{n+1}$, employing half-step fluxes $\boldsymbol{F}^{n+1/2}$ and explicit source terms $\boldsymbol{S}^{n+1/2}_{E}$, together with end-of-step source terms $\boldsymbol{S}^{n+1}_{I}$ and $\boldsymbol{S}^{n+1}_{T}$, to obtain the updated primitives $\boldsymbol{P}^{n+1}$.

We validate the EGRMHD implementation in \kharma using a suite of test problems detailed in \cite{chandra_grim_2017}. Figure \ref{fig:emhd_convergence_kharma} shows convergence results from a representative subset of these tests, each probing distinct aspects of the numerical algorithm described in the previous section. The norm of the $L_{1}$ error decreases with resolution at the anticipated order of convergence.

\section{Comparison Between SANE and MAD Extended GRMHD Simulations}
\label{appendix:sane_vs_mad_egrmhd}

\begin{figure*}
\centering
\includegraphics[,width=\linewidth]{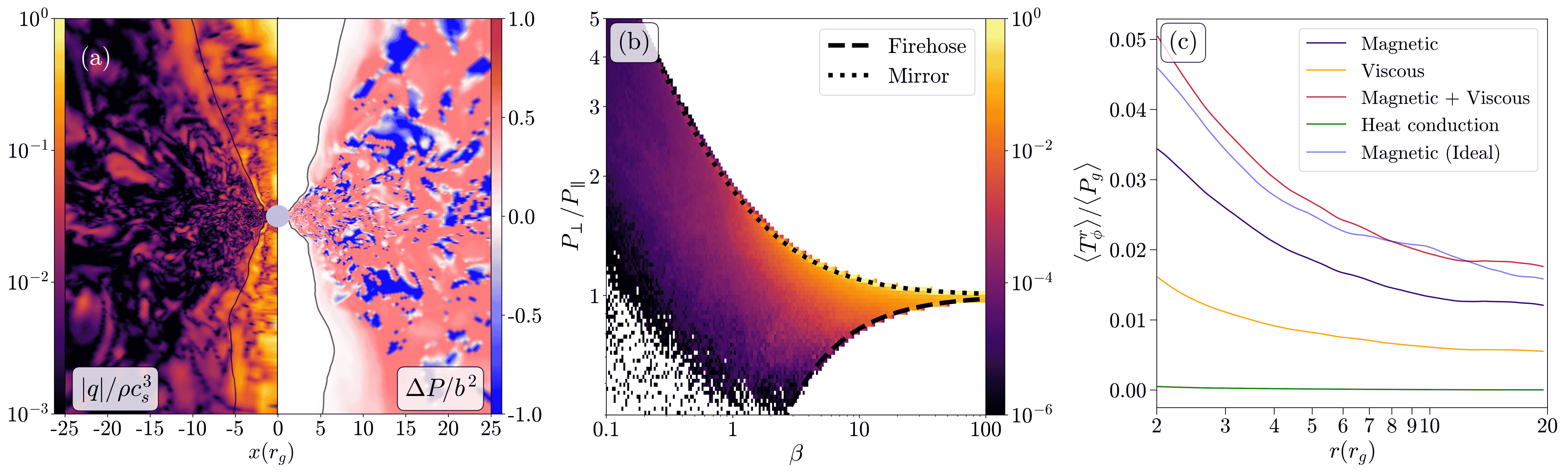}
\caption{GRMHD quantities from the SANE $a_{*} = 15/16$ simulation. (a) Poloidal slices of the dissipative variables: heat flux normalized by the free-streaming value (left) and pressure anisotropy normalized by magnetic energy density (right). The black contour marks $\sigma = 1$. (b) Mass-weighted distribution of pressure anisotropy as a function of $\beta$ within $r \leq 20r_{\text{g}}$. (c) Time-averaged radial profile of angular momentum transport in the disk.}\label{fig:emhd_snapshot_sane}
\end{figure*}

The main text primarily focuses on magnetically dominated flows.  Here we examine SANE simulations and compare the evolution of $q$ and $\Delta P$ with that observed in MAD accretion.

\begin{figure*}
\centering
\includegraphics[,width=\linewidth]{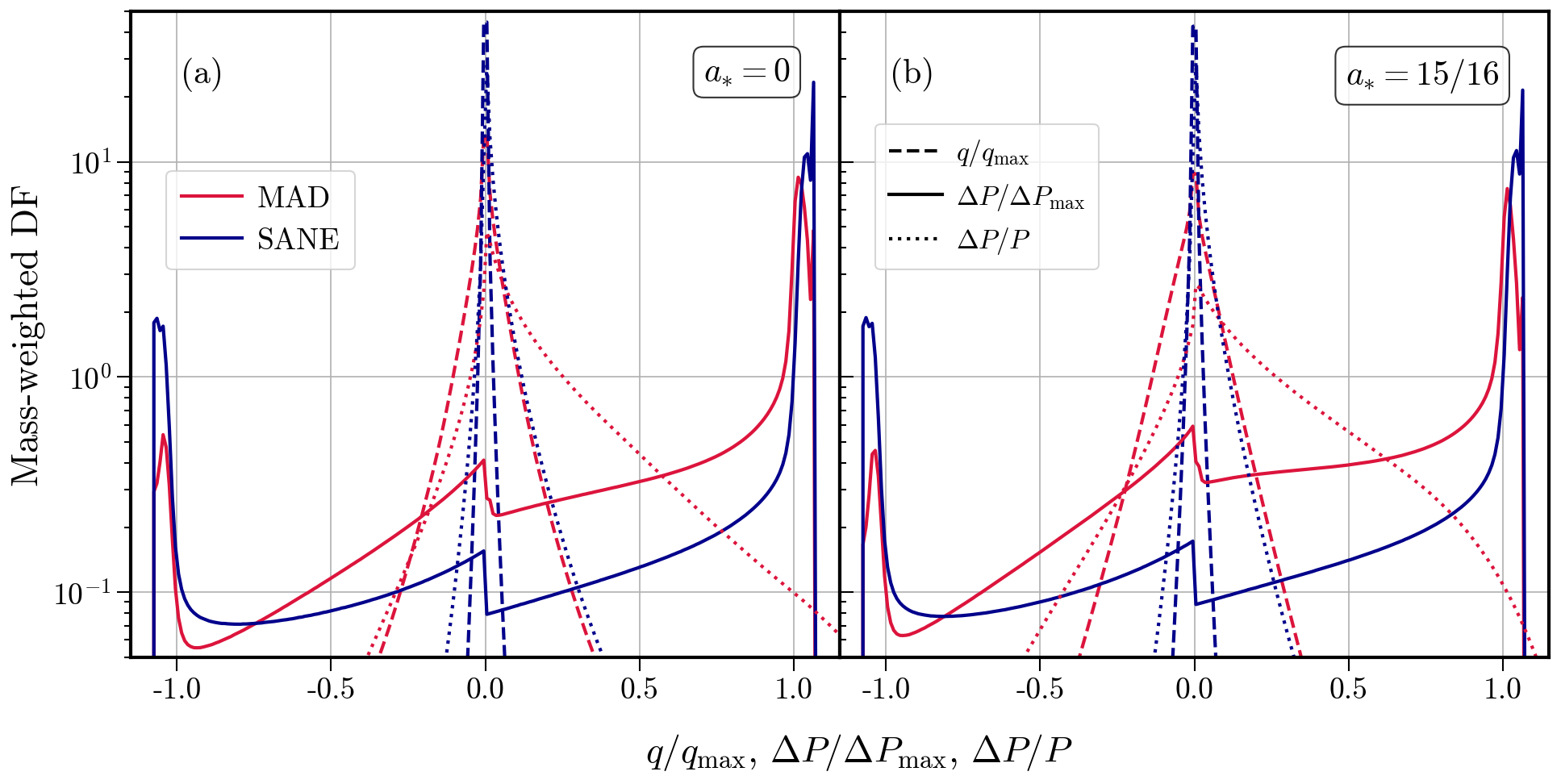}
\caption{Time-averaged, mass-weighted distribution functions of $q/q_{\text{max}}$, $\Delta P/\Delta P_{\text{max}}$ and $\Delta P / P$ for (a) $a_{*}=0$ and (b) $a_{*}=15/16$ simulations. $\Delta P_{\text{max}}=\Delta P_{\text{mirror}}$ if $\Delta P>0$ and $\Delta P_{\text{max}}=\Delta P_{\text{firehose}}$ if  $\Delta P<0$.}
\label{fig:emhd_qdP_distribution}
\end{figure*}

Panel (a) in Figure \ref{fig:emhd_snapshot_sane} presents a poloidal snapshot from the SANE $a_{*}=15/16$ simulation. The heat flux is negligible within the disk $q\lesssim0.01q_{\text{max}}$, and is an order of magnitude lower than what is observed in MAD flows. This is evident in mass-weighted distribution profiles of $q/q_{\text{max}}$ plotted in Figure \ref{fig:emhd_qdP_distribution} (dashed lines)---the normalized heat flux has a smaller spread about zero for SANE models. This can be explained by noticing that a larger fraction of the disk mass saturates at one of the instability thresholds ($\sim65\%$ attains $\Delta P_{\text{mirror}}$ and $\sim10\%$ reaches $\Delta P_{\text{firehose}}$) compared to MAD simulations (see Figure \ref{fig:emhd_snapshot_sane} panel (b) and cf. Figure 1 (c) in main text). This leads to a suppression of $\tau_{\text{R}}$ and, hence, the target value $q_{0}$ because $q_{0}\propto\chi\propto\tau_{\text{R}}$. Consequently, viscous stresses in SANE simulations contribute nearly half as much as magnetic stresses to angular momentum transport in the disk (Figure \ref{fig:emhd_snapshot_sane}c). Although $\langle\Delta P\rangle / \langle b^{2}\rangle$ is greater in the SANE case, the accretion disks are weakly magnetized, $\beta\gtrsim10$, and as a result the deviations of fluid pressure from the ideal gas value are more strongly constrained by plasma instabilities. We see in Figure \ref{fig:emhd_qdP_distribution} that $\Delta P/P\lesssim 0.3$ in SANE models (dotted dark blue lines) while $\Delta P/P$ can exceed unity in MAD flows.

\begin{figure*}
\centering
\includegraphics[,width=\linewidth]{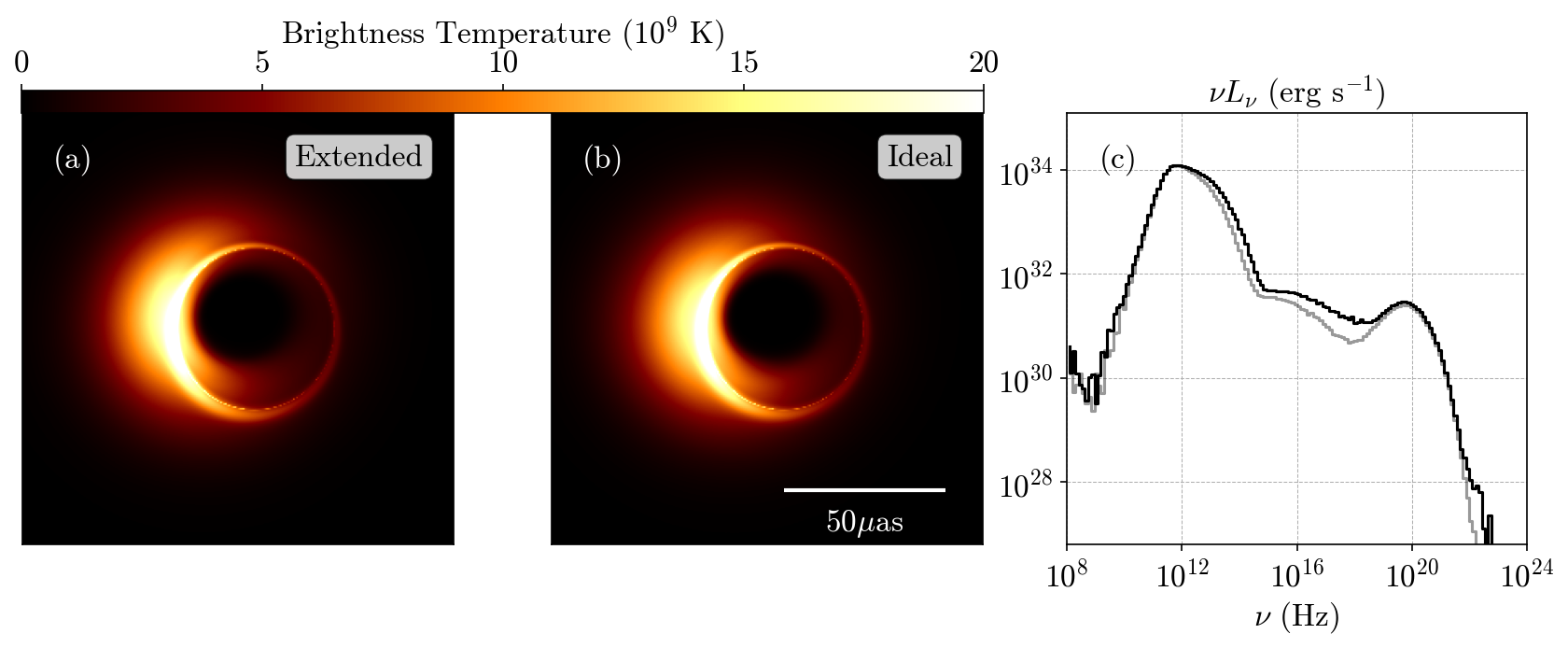}
\caption{Electromagnetic observables for SANE $a_{*}=15/16$ simulation with $R_{\text{high}}=10$ at a viewing angle of $10\degree$. (a) and (b) Time-averaged 230 GHz total intensity images for the Extended and Ideal GRMHD simulations, respectively. (c) Comparison of time-averaged spectra for the simulations shown in (a) and (b).}
\label{fig:grrt_comparison_sane}
\end{figure*}

Finally, Figure \ref{fig:grrt_comparison_sane} presents time-averaged electromagnetic observables for an EGRMHD SANE model and its ideal counterpart. As in the MAD case, the synthetic images of both plasma models are nearly identical. The SEDs also show close agreement, with the EGRMHD model exhibiting slightly higher power at $\nu\sim10^{18}$ Hz, primarily due to increased Compton upscattering.

\bibliography{main}
\bibliographystyle{aasjournal}

\end{document}